\documentclass[12pt]{article}
\usepackage[T1]{fontenc}
\usepackage{marginnote}
\usepackage[english]{babel}
\usepackage{graphicx}
\usepackage{natbib}
\usepackage[textfont=sc,labelfont=bf]{caption}
\usepackage[blocks]{authblk}
\usepackage{setspace}
\usepackage{multirow}
\usepackage{amsthm}
\usepackage{amsmath}
\usepackage{amsfonts}
\usepackage{amssymb}
\usepackage{pgf} 
\usepackage{bm}
\usepackage{paralist}
\usepackage{multirow}

\newtheorem{ass}{Assumption}

\newcommand{\beq}{\begin{equation}}
\newcommand{\eeq}{\end{equation}}
\newcommand{\beqa}{\begin{eqnarray}}
\newcommand{\eeqa}{\end{eqnarray}}
\newcommand{\beqan}{\begin{eqnarray*}}
\newcommand{\eeqan}{\end{eqnarray*}}

\newcommand{\beqy}{\begin{eqnarray}}
\newcommand{\eeqy}{\end{eqnarray}}

{
}

\usepackage{graphicx}

\def\d{\displaystyle}
\numberwithin{equation}{section}
\definecolor{amaranth}{rgb}{0.9, 0.17, 0.31}
\definecolor{amber}{rgb}{1.0, 0.75, 0.0}
\definecolor{brickred}{rgb}{0.8, 0.25, 0.33}
\definecolor{brilliantrose}{rgb}{1.0, 0.33, 0.64}
\definecolor{byzantine}{rgb}{0.74, 0.2, 0.64}
\definecolor{carmine}{rgb}{0.59, 0.0, 0.09}
\definecolor{frenchblue}{rgb}{0.0, 0.45, 0.73}
\definecolor{antiquefuchsia}{rgb}{0.57, 0.36, 0.51}
\definecolor{chocolate(web)}{rgb}{0.82, 0.41, 0.12}
\definecolor{ceruleanblue}{rgb}{0.16, 0.32, 0.75}
\definecolor{green(colorwheel)(x11green)}{rgb}{0.0, 1.0, 0.0}
\definecolor{dollarbill}{rgb}{0.52, 0.73, 0.4}
\definecolor{maroon}{rgb}{128,0,0}
\definecolor{auburn}{rgb}{0.43, 0.21, 0.1}
\definecolor{awesome}{rgb}{1.0, 0.13, 0.32}
\definecolor{cardinal}{rgb}{0.0, 0.0, 1.0}
\definecolor{arsenic}{rgb}{0.23, 0.27, 0.29}
\definecolor{amber(sae/ece)}{rgb}{1.0, 0.49, 0.0}
\definecolor{cerulean}{rgb}{0.0, 0.48, 0.65}
\definecolor{ashgrey}{rgb}{0.7, 0.75, 0.71}
\definecolor{battleshipgrey}{rgb}{0.52, 0.52, 0.51}
\definecolor{brown(traditional)}{rgb}{0.59, 0.29, 0.0}
\definecolor{bulgarianrose}{rgb}{0.28, 0.02, 0.03}
\definecolor{cardinal}{rgb}{0.77, 0.12, 0.23}
\definecolor{lava}{rgb}{0.81, 0.06, 0.13}
\definecolor{persimmon}{rgb}{0.93, 0.35, 0.0}

\begin{document}

\begin{titlepage}
%\pagenumbering{gobble}
\title{\vskip-.8truecm High-Dimensional Dynamic Factor Models: A Selective Survey and Lines of Future Research}
\author{Marco Lippi\thanks{E-mail: mlippi.eief@gmail.com.}  \\ \vskip-.4truecm  
Einaudi Institute for Economics and Finance \and

Manfred Deistler\thanks{E-mail: manfred.deistler@tuwien.ac.at. }\\
\vskip-.4truecm
Technische Universit\"at Wien and Wirtschaftsuniversit\"at Wien\and

Brian Anderson\thanks{E-mail: Brian.Anderson@anu.edu.au. School of Engineering, Australian National University, Acton, ACT 2601, Australia.} \\ %\vskip-.4truecm  
Australian National University
}

\date{\today}
\maketitle	
\begin{abstract} 
\noindent High-Dimensional Dynamic Factor Models are presented in detail: The main assumptions and their motivation, main results,  illustrations by means of elementary examples. In particular, the role of singular ARMA models in the theory and applications 
of High-Dimensional Dynamic Factor Models is discussed.
The emphasis of the paper is on model classes and their structure theory, rather than on estimation in the narrow sense. Our aim is not a comprehensive survey. Rather  we try to  point out  promising lines of research and applications that have not yet been sufficiently developed.
\end{abstract}
\thispagestyle{empty}
\end{titlepage}

\section{Introduction}\label{sec:uno}

Analysis and forecasting of high-dimensional time series recently  has attracted substantial interest, see \cite{hallin20}.  However,  ``classical'' multivariate time-series  models such as autoregressive models suffer from the so-called ``curse of dimensionality'': Unless additional restrictions are imposed,  the parameter spaces grow with the square of the dimension of the time series, $N$ say; thus in many cases, even for moderate $N$, the sample size available is not sufficient to guarantee reliable parameter estimation. This is true in particular for macroeconomic applications. In this case, on the one hand, the interaction of several variables,  disaggregated  series in particular, may be important for modeling and forecasting. On the other hand, sample size  of macroeconomic time series  is in many cases rather limited,   e.g.  because of structural changes in the underlying economies.

High-dimensional dynamic
factor models are one way to overcome this curse of dimensionality. 
Factor models for the case of i.i.d. observations have a long history, dating back to \cite{spearman04}  and  \cite{burt09}. Factor models in time-series context have been proposed much later, in particular by 
\cite{geweke77} and \cite{SargentSims}, \cite{WatsonEngle83},  \cite{QuahSargent93}.  The basic idea is to separately model the comovement between the variables  on the one hand, and  the individual, or { idiosyncratic} movement on the other hand.  Equivalently, the variables are decomposed as the sum of a {\it latent variable}, or {\it common component},  and an {\it idiosyncratic} component.

The above factor models are ``exact'' in the sense that the idiosyncratic components are assumed to be cross-sectionally uncorrelated at all leads and lags. Inspired by the idea of risk diversification,    \cite{Chamberlain} and
\cite{chamberlainrotshild83} introduced the notion of  ``approximate''   factor models, which are high-dimensional (potentially infinite-dimensional) models  where  the idiosyncratic terms are allowed to be cross-sectionally  dependent, although in a weak sense. This idea has been extended to the linear dynamic case by 
\cite{forni_generalized_2000}, \cite{fornilippi01}, \cite{stockwatson02JASA,stockwatson02JBES}, \cite{baing02},  \cite{bai03}, leading to the class of  linear High-Dimensional Dynamic Factor Models, Dynamic Factor Models for short  (DFM). 

The present paper does not contain a comprehensive survey. We go over the main assumptions and results with the purpose of pointing out lines of research that have not yet been sufficiently developed. 

In 
 Section \ref{sec:due} we describe the model class of DFMs, thus the decomposition of the observable $N$-dimensional  vector $y^N_t$ into common and idioyncratic components, $\chi^N_t$ and $\xi^N_t$ respectively.  We assume that  $\chi_t^N$ has rational spectral density $f^N_\chi$. The rank  $q$ of $f^N_\chi$ does not depend on $N$, for $N$ sufficiently  large. The state dimension of minimal, stable and miniphase state space realization of $f^N_\chi$ is independent of $N$.  (This can be rephrased as saying, that there is a finite number of  ``static'' factors, see Section~\ref{sec:tre}.)
 
  In Section \ref{sec:tre} we discuss the model for the process of the common components.  Precisely, the common components are represented as linear combinations of a finite-dimensional vector process, whose coordinates are called the (minimal) static factors. The latter are modeled as an ARMA system driven by a vector white noise whose coordinates are called the dynamic factors.
  Here  we introduce the so-called 
 singular   ARMA  (or state-space) systems.  These generate $N$-dimensional stationary processes with rational spectral density, whose rank is less than $N$: we call them singular\footnote{Not to be confused with singular processes in the sense of Kolmogorov.}. We argue that, under our assumptions,  this  is obviously the case for 
 the common-component vector, and is very likely for the factor vector.  On  the other hand, it has been shown that singular ARMA processes can (generically) be modeled as finite
 AR's, see Section \ref{sec:tre.tre}, which implies a most important simplification in the modeling of common components and static factors.

 %This is done in two steps: in a first step (minimal) static factors are extracted and in a second step the dynamics of the minimal static factors are modeled by ARMA or AR systems. 
 
In Section \ref{Denoising} we describe   techniques  to obtain the common components and factors from the observable vector $y^N_t$,  for $N$ and $T$ (the number of observations for each time series) tending to infinity. In Section  \ref{subsec:PC} we  show how principal components (PCA) can be  used. The underlying estimation procedure is to estimate the static factors by PCA in a first step and then to estimate an ARMA or AR model in order to describe the dynamics of the static factors. An alternative approach, see Section \ref{sec:quattro.due},
is to assume a dynamic factor model with autoregressive static factors and cross-sectionally uncorrelated idiosyncratic components, thus an exact factor model, and to put this in a state-space framework. In this framework (once the integer specification parameters have been fixed) an EM algorithm with the  E-step based on  Kalman filtering can be applied.  That an exact factor model can be used to estimate a DFM has been shown in \cite{DGRqml}, see \ref{sec:quattro.due}.

In the case where no $N$-independent static factors exist, 
frequency-domain methods, as described in Section 
\ref{appiccicata}, may be applied.

In Section 
\ref{subsec:quattro.uno} we  present some 
applications of the results on   singular ARMA models to  empirical macroeconomic analysis, to the so-called fundamentalness problem in particular.  Some results on cointegration for  singualar ARIMAs are presented in Section \ref{NS}.

\section{High-Dimensional Dynamic Factor Models (DFM). The Model Class}\label{sec:due}
%\subsection{Assumptions}\label{sec:due.uno}

The  basic idea is to represent the $N$-dimensional observation vector at time $t\in \mathbb Z$, $y^N_t$ say, as 
\begin{equation} \label{eq:due.uno} y^N_t=\chi^N_t+ \xi^N_t, \end{equation}
where $(\chi^N_t\;|\; t\in \mathbb Z)$ and $( \xi^N_t\;|\; t\in \mathbb Z)$ are  the $N$-dimensional processes of common and idiosyncratic components respectively. The one-dimensional { processes  $(\chi_{it}, \ t\in \mathbb Z) $ are strongly dependent 
across  the index $i$,   whereas the processes  $(\xi_{it}, \ t\in \mathbb Z)$ are  weakly dependent. The precise meaning of strong and weak dependence is given below.}

Throughout, except for  Section \ref{NS}, we assume that  $(\chi^N_t)$ and $(\xi^N_t)$ are wide-sense stationary.
%{\color{cardinal} with absolutely summable covariances and thus with a  spectral density defined and continuous everywhere in $[-\pi,\ \pi]$.} 
In addition, throughout we assume 
\begin{align}
{\rm E} \chi^N_t ={\rm E } \xi^N_t&=0\ \ \  \forall t\label{eq:due.due}\\
{\rm E}\chi^N_t\, {\xi^N_s}'&=0\ \ \ \forall t,s,
\label{eq:due.tre}\end{align}
{and that the spectral densities of ($\chi^N_t$) and ($\xi^N_t$) exist.}  As a consequence,   ($y^N_t$)  is stationary and has a spectral density, which,
 using an obvious notation, is: 
\begin{equation} f_y^N(\lambda) = f_{\chi}^N (\lambda) + f^N_\xi(\lambda)\label{eq:due.quattro}, \ \ \  \lambda\in [-\pi,\ \pi].\end{equation}  

% In the standard version of the GDFM, to be defined below,				%	 the latent variables $\chi_{it}$  can be represented:  (i)~as %linear combinations,   { or limits of such linear combinations}, of   %{\it dynamic  factors} and their lags, see Section \ref{subsec:tre.uno}, (ii)~as linear
%combinations of {\it static factors}, see  Section \ref{subsec:tre.due}.  

Throughout, $z$ is used for a complex variable as well as for the backward shift  on $\mathbb Z$.

The following assumptions constitute the class of DFM's considered in the present paper (we follow here \cite{andersondeistlersingular}): 
%(nonstationary models are considered in   Section %\ref{NS}: 

\begin{ass} \label{ass:uno}  For all $N$, $f^N_{\chi}$  is a rational spectral density. \end{ass}

 An obvious consequence of Assumption \ref{ass:uno} is that  $f_{\chi}^N$ has constant normal  rank,  i.e. has the same rank almost everywhere
 on $[-\pi,\ \pi]$. 
 
Here, for asymptotic analysis not only the sample size $T$, but also the cross-sectional dimension $N$ is tending to infinity---this  has an empirical motivation in the study of { high-dimensional  vector  time series, i.e. vector time series  whose dimension  is allowed to be close { to} or even higher than the sample size.    Thus the underlying process considered is a double-indexed stochastic process $(y_{it}\; |\; i\in \mathbb N,\ t\in \mathbb Z)$, corresponding, as $N$ varies, to a nested sequence of models \eqref{eq:due.uno}, in the sense that $y_{it}, $ $\chi_{it},$ and $\xi_{it}$ do not depend  on $N$, for $i\leq N$.

\begin{ass}\label{ass:due}  {We suppose that there exists $N_0\geq q$  such that, from}  $N_0$ onwards,
the rank of $f^N_{\chi} $ is independent of $N$.   Such rank is denoted  by $q$.
\end{ass}

{ As is well known, see e.g. \cite{deistler12}, every rational spectral density can be described by an ARMA or alternatively a state space system.}

\begin{ass}\label{ass:tre} The state dimension of a minimal, stable and miniphase state space realization corresponding to $f_{\chi}^N$ is independent of $N$ {  from a certain $N_1$ onwards.}   Such state dimension is denoted by $n$.
\end{ass}

Of course, without loss of generality, we can assume $N_0=N_1$.
%{\color{cardinal}  \begin{ass}\label{ass:treaggiu}For $N\geq N_0$, the dimension of the space 
%spanned by $\chi^N_t$, for any given $t$   is independent of $N$. Such dimension is denoted by $r$.\end{ass}
%
%It is fairly easy to show, see Sections \ref{subsec:tre.uno} and \ref{subsec:tre.due}, that Assumptions \ref{ass:tre} and \ref{ass:treaggiu} are equivalent.}
Next we define weak and strong  dependence.
We use   $\omega ^N_{\xi,s}(\lambda)$, $\omega^N_{\chi,s}(\lambda)$, $\omega _{y,s}^N(\lambda)$ to denote the $s$-th largest eigenvalue of the Hermitian matrix $f^N_\xi(\lambda)$,  $f^N_{\chi}(\lambda)$, $f^N_{ y}(\lambda)$, respectively. 

\begin{ass} \label{ass:quattro} Weak cross-sectional dependence of $(\xi^N_t)$. { The eigenvalues  $\omega^N_{\xi,1}(\lambda)$ are uniformly bounded,} i.e. there exists $B>0$ such that $\omega^N_{\xi,1}(\lambda)<B$  for all $N$ and $\lambda$.
\end{ass}

\begin{ass} \label{ass:cinque} Strong cross-sectional dependence of $(\chi^N_t)$. The  eigenvalues $\omega^N_{\chi,s}(\lambda)$, $s=1,\ldots,q$, diverge as $N\to \infty$, $\lambda$ almost everywhere in $[-\pi,\ \pi]$. 
\end{ass}

Note that  Assumptions   \ref{ass:quattro} and \ref{ass:cinque} place restrictions on the {\it cross-dependence} of the idiosyncratic and common component respectively, not on their autocorrelations. 
%In the case where $(\chi^N_t)$ and $(\xi^N_t)$ are white-noise processes, so that  $ f^N_{\chi}$ and $f_\xi^N$ are constant as functions of $\lambda$,  for all $N$, \eqref{eq:due.uno} is  called a  large-dimensional {\it static} factor model.
Note also that the assumptions defining  the common and idiosyncratic terms  are asymptotic, for the number $N$ of observable variables tending to infinity, so that in empirical application  the observable vector $y^N_t$ is supposed to be high dimensional.
 
{ \cite{fornilippi01} prove that   representation \eqref{eq:due.uno} is identified. More precisely, 
if $\tilde\chi^N_t$, $\tilde\xi^N_t$ and the integer $\tilde q$ fulfill
$$ y^N_t= \tilde \chi^N_t+ \tilde\xi^N_t,\ \  \hbox {for all $N$}, $$
\eqref{eq:due.due}, \eqref{eq:due.tre}  and Assumptions  \ref{ass:uno}, \ref{ass:due}, \ref{ass:quattro} and  \ref{ass:cinque}, then $\tilde q=q$, $\tilde\chi^N_t= \chi^N_t$,
$\tilde\xi^N_t=\xi^N_t$.}

DFM's generalize exact dynamic factor models,  considered in \cite{geweke77}, \cite{SargentSims}, where  $N$ is fixed and$f_\xi^N(\lambda) $ is diagonal, i.e.  the one-dimensional   idiosyncratic processes are mutually uncorrelated at any lead and lag.   Exact dynamic factor models are rather restrictive, in  that the class of spectral densities corresponding to them is  restricted, in particular  if $q$ is small relative  to $N$ (which is the most important case), see 
\cite{scherdeist98}. On the other hand, for $q$ small enough in relation to $N$, identifiability results are available for { fixed $N$}, see again \cite{scherdeist98}. 

%The idea to generalize  static factor models  is due %to \cite{Chamberlain} and %\cite{chamberlainrotshild83} who developed a theory %by allowing for weak dependence of the noise %component in a high-dimensional context.

In general,  several interpretations  for DFM's are possible. E.g., the components $\xi_t^N$  may be interpreted as  ``measurement'' error and $\chi^N_t$ as unobserved ``true'' variables; this is in line with errors-in-variables 
models, which, from an abstract point of view, are the same as factor models.  An alternative interpretation of \eqref{eq:due.uno} is the decomposition of observations into a part $(\chi^N_t$) representing the comovements
between the variables (for instance representing the ``market'' { effect on all stock prices)} and individual movements ($\xi^N_t$) { (representing the firm's specific effect)}. The components $\xi_{it}$ are usually called the  {\it idiosyncratic components}, while the latent variables $\chi_{it}$ are referred to as the {\it common components}.

It is easy to see that,  given a bijective map $g:\mathbb N\to \mathbb N$, if Assumption \ref{ass:cinque} holds for  the  process $(\chi_{it}\; |\; i\in \mathbb N,\ t\in \mathbb Z)$, then it holds for $(\chi_{g(i),t}\; |\; i\in \mathbb N,\ t\in \mathbb Z)$, because   the vector 
$\chi^N_t$ is nested in  the vector $ \chi^{g,M} _t= (\chi_{g(1),t}\ \cdots\ \chi_{g(M),t})$ for some $M$.  As a consequence, $\omega ^N_{\chi, s}(\lambda)\leq \omega ^{g,M}_{\chi,s}(\lambda)$, see e.g. \cite{fornilippi01}, Fact M, (b), p. 1121. 
Thus Assumption \ref{ass:cinque} holds irrespective of the order of
the variables $y_{it}$ (and thus $\chi_{it})$. 

However, the speed  of divergence of the eigenvalues $\omega^N_{\chi,s}(\lambda)$, $s=1,\ldots,q$,
depends on that order.
A simple example is the following.  Let  $(a_s,\ s\in \mathbb N)$ be a square summable 
sequence of real numbers and $(j_s,\ s\in \mathbb N)$  a sequence of positive integers. Then let 
$$ \chi_{it} = b_i v_t,$$
where $v_t$ is a unit-variance scalar white noise and the coefficients $b_i$ are the following:
\begin{equation}\label{eq:due.aggiu} 1,\ a_1,\ \ldots a_{j_1},\ 1,\ a_{j_1+1},\ \ldots \ a_{j_1+j_2}, \ 1,\   a_{j_1+j_2+1},\ \ldots\end{equation}
In this case we have 
$$\omega^N _{\chi,1}(\lambda) =\sum _{h=1}^N b_h^2.$$  
In particular, if $N= K + j_1+\ldots + j_K$, 
$$ \omega^N _{\chi,1}(\lambda)= K + a_1^2+ a_2^2 + \cdots + a_{j_1+\cdots +j_K}^2.$$
Assuming that $j_s=1$ for all $s$, we see that, for $N$ odd,
$$\lim _{N\to \infty}\frac{\omega^N _{\chi,1}(\lambda)}{N} =\lim _{N\to \infty}\frac{K+a_1^2+\cdots + a_K^2}{2K}=
 \frac{1}{2}.$$
On the other hand, assuming that 
$j_s=s$,  we obtain a reordering of the coefficients $b_i$,  with ones  appearing at linearly increasing intervals. It is easy to see that in this case
the eigenvalue $\omega^N _{\chi,1}(\lambda)$ grows asymptotically 
at  speed $N^{1/2}$.  %In particular, if $a_s=0$ for all $s$, we see that the consequence of  sparsity of non-zero coefficients 
%depends on the order in which the variables are  added to the model. 
{ DFM's in which the eigenvalues are assumed to diverge 
with rates $N^\alpha$, $0<\alpha<1$,  known as  models with ``weak factors'',
necessarily rely on a particular assumption on the order of the variables, 
see \cite{onatskiweak} and related literature.}

\section{Modeling the Latent Process}\label{sec:tre}
\def\d{\displaystyle}
\subsection{Singular {ARMA} and state space systems}\label{subsec:tre.uno}

By Assumptions \ref{ass:uno}  and  \ref{ass:due}, the latent process $(\chi^N_t$)  has a rational and, for { $N>N_0\geq  q$,} singular spectral density. By its rationality,   it can be represented by a singular  ARMA or state space system, i.e. an ARMA or state space system with a singular innovation variance. For regular ARMA or state space systems, i.e. systems  whose output has an a.e. nonsingular spectral density, we refer to 
\cite{deistler12}, Deistler and Scherrer (2019); for the singular case see Section \ref{sec:tre.tre}.

Hereafter, unless strictly necessary, we omit the superscript $N$.
For the ARMA system,   assuming that $N\geq N_0$,  we have
\begin{equation}\label{eq:tre.uno} P(z) \chi_t=Q(z) v _t, \ \ \
P(z) = P_0-\sum _{j=1}^S P_jz^j,\ \ \  Q(z) =\sum^{S'} _{j=0} Q_jz^j ,\end{equation}
where $v_t$ is an orthonormal $q$-dimensional white noise, 
with $ P_j\in \mathbb R^{N\times N}$,  $Q_j\in \mathbb R^{N\times q}$.   Moreover,
\begin{equation} \label{eq:tre.due} \det P(z) \neq 0 \ \ {\rm for}\ \ |z|\leq 1,\end{equation}
which is the  stability condition, and 
\begin{equation}\label{eq:tre.aggiu} {\rm rank } \, Q(z) = q, \hbox{ \ \ for  $|z|\leq 1$}, \end{equation}
which is   the strict miniphase condition. \def\Es{{\rm E}}

  Thus  the steady state solution of \eqref{eq:tre.uno} is 
  \begin{equation} \label{eq:tre.tre} 
  \chi_t =P^{-1} (z) Q(z) v_t=  k(z)v_t{ = \sum_{j=0} ^\infty k_j v_{t-j}}.\end{equation}

Due to the stability and the miniphase conditions, \eqref{eq:tre.tre}  is a Wold representation, { $v_t $ are innovations  and the one-step ahead prediction error for $\chi_t$, given its past $\chi_s$, $s<t$, is $k_0 v_t$, see below.}
%If the variance matrix of the innovations $(\epsilon_t)$ is singular then the ARMA system \eqref{eq:tre.uno} (or
%$(a(z)\ \tilde b(z))$ respectively) is called {\it singular}.   For singular ARMA systems see e.g. Anderson et al. (2012), Chen et al. (????) and  Deistler (2019).
The spectral density of $(\chi_t)$  is of the form 
\begin{equation} f_{\chi}(\lambda) = (2\pi)^{-1} k(e^{-i\lambda} )k(e^{-i\lambda})^*\label{eq:tre.cinque}\end{equation}
where $*$ denotes the Hermite conjugation.  $f_{\chi}(\lambda)$ has rank $q$ for all $\lambda \in [-\pi,\ \pi]$.  Given $f_{\chi}$,   under  conditions  \eqref{eq:tre.due} and \eqref{eq:tre.aggiu}, }
$k(z)$ is unique up to postmultiplication by (constant) orthogonal matrices.  Moreover:

{
\begin{ass}\label{ass:febbraio}  We suppose that
 $P_0=I_N$ and that $P(z) $ and $ Q(z)$ are left coprime  \cite[see][ p.41]{deistler12}), i.e.  the matrix
$(P(z)\  Q(z))$ has rank $N$ for all $z\in \mathbb C$.
\end{ass}
}
%{ By Assumption \ref{ass:tre}, for $N>N_0$ the space spanned by $\chi^N _{t-k}$, $k\geq 0$, 
%is equal to the space spanned by $\chi^{N_0+1}_{t-k}$.  This implies, by Lemma 1 in 
%\cite{FHLZ14},   that for $N>N_0$  the $q$-dimensional white noise $v_t$ is independent 
%of $N$, and that the matrices $a(z)$ and  $b(z)$ are nested.}

Alternatively, 
by Assumptions \ref{ass:due} and \ref{ass:tre}, for $N\geq N_0$,    to $f_{\chi} (\lambda)$ there corresponds a minimal  state-space realization 
\begin{align} \label{eq:tre.sei} x_{t+1} &= F x_t + G w_{t+1} \\ \label{eq:tre.sette} \chi_t &= H x_t,\end{align}
where $x_t$ is an $n$-dimensional state, $w_t$ is a $q$-dimensional 
orthonormal white noise, { where $n$ and $q$ are independent of $N$, }  $F\in \mathbb R^{n\times n} $, $G\in \mathbb R^{n\times q}$,  $H\in \mathbb R^{N\times n}$ are parameter matrices.   The stability condition is
\begin{equation}\label{eq:gennaio1}\rho (F) <1,   \end{equation}
where $\rho(F) $ denotes the spectral radius of $F$, and
\begin{equation}\label{eq:gennaio2} {\rm rank}\, \begin{pmatrix}  I-Fz & -G \\ H & \phantom{-}0\\ \end{pmatrix}=n+q,\ \  {\rm for }\ \ |z|\geq 1
\end{equation}
is the strict miniphase condition.
%{\color{frenchblue}  MANFRED, I DO NOT UNDERSTAND THE MINIPHASE CONDITION.}

Under  conditions \eqref{eq:gennaio1} and \eqref{eq:gennaio2}, $w_t$ is an innovation for $\chi_t$. Moreover, 
setting
$$ K(z) = H(z^{-1}I-F)^{-1} G,$$
we have 
\begin{equation} \label{eq:tre.otto} \chi_t = H(I-Fz)^{-1} G w_t=K(z)w_t.\end{equation}
Under our assumptions, the innovations and the transfer function  are unique up to  premultiplication and postmultiplication  by orthogonal matrices, respectively, so that we can assume with no loss of generality  that $w_t=v_t$  and $K(z)=k(z)$, where $v_t$ and $k(z)$ are defined in \eqref{eq:tre.uno} and \eqref{eq:tre.tre} respectively.

{  Representation \eqref{eq:tre.sei}--\eqref{eq:tre.sette} implies that, for   $N\geq N_0$,  the dimension of the space spanned  by $\chi _{1t},\ \chi_{2t},\ \ldots,\ \chi_{Nt}$, call it ${\cal S}^N_t$,  for any given $t$,  does not exceed  $n$.     Therefore   there exists $\tilde N\geq N_0$ and $r\leq n$ such that 
for $N\geq \tilde N$,  
\begin{equation}\label{eq:3.nuova}  {\rm dim}({\cal S}^N_t)=r. \end{equation} 
 With no loss of generality we can assume that   $\tilde N=N_0$.  By \eqref{eq:3.nuova}, 
  $\chi_{N_0+k,t}$, $k>0$, is a linear combination of $\chi_{1t},\ \chi_{2t},\ \ldots,\ \chi_{N_0,t}$, so that 
$\chi_{N_0+k,t}= H_i x_t$, $H_i\in \mathbb R^{1\times n}$, where $x_t$ is the state vector of the minimal state-space realization for $N=N_0$.  Thus, 
representation \eqref{eq:tre.sei}--\eqref{eq:tre.sette} holds for all $N\geq N_0$ with the same $x_t$, $w_t$, $F$ and $G$, and nested matrices $H$.  Otherwise stated,  \eqref{eq:tre.sei}--\eqref{eq:tre.sette} hold with 
\eqref{eq:tre.sette} replaced by $\chi_{it}=H_i x_t$, where $H_i$ is  the $i$-row of the 
matrix $H$.
}

{ Note  that there is a Hilbert space (in the Hilbert space $L_2$ of square integrable random variables) construction of a state-space system, obtained  by projecting all future values of  the output process $(\chi_t)$ on the Hilbert space spanned by its past  (see e.g. Akaike 1974). Then, by the rationality of the spectral density, the space spanned by these projections is finite dimensional  and every basis is a minimal state, see e.g. \cite{deistler12}, \cite{deischerrer18}. We will call this the Kalman-Akaike realization of a state-space system.  

The state space system \eqref{eq:tre.sei}--\eqref{eq:tre.sette} corresponds to a definition of past and future, respectively, as $\chi_s$, $s\leq t$ and $\chi_s$, $s\geq t$. If we, however, define the past as $\chi_s$, $s<t$, the  Kalman-Akaike realization leads to a system of the form
\begin{align} \label{eq:tre.seisei}\bar  x_{t+1} &= \bar F \bar x_t + \bar G v_{t} \\ \label{eq:tre.settesette} \chi_t &=\bar  H \bar x_t+ B_0 v_t\end{align}
and the corresponding representation  of the transfer function is 
\begin{equation} \label{eq:ottootto} \bar H (Iz^{-1}-\bar F )^{-1} \bar G +B_0.\end{equation}
Note that  now we impose a stability and  a minimum phase condition analogous to  \eqref{eq:gennaio1} and  \eqref{eq:gennaio2}
respectively,  for $(\bar F,\bar G,\bar H, B_0)$. Moreover, $\bar H\bar x_t$ is the one-step ahead forecast for $\chi_t$ and 
$B_0 v_t$ the corresponding one-step  ahead prediction error.

}
%, the dimension of the state $x_t$ does not change as $N$ increases.
%Thus \eqref{eq:tre.sette} implies that
%$\chi_{N_0+k,t}$, $k\geq 2$,  is linearly dependent on $\chi_{1t},\ldots,\chi_{N_0+1,t}$, so that the same $x_t$, $F$, $H$ and innovations  $(v_t)$  can be taken   for all $N> N_0$.}
%

%Under stability and the miniphase condition, $w_t$ is an innovation for $\chi_t$. Moreover, 
%setting
%$$ K(z) = H(1-Fz)^{-1} G,$$
%we have 
%\begin{equation} \label{eq:tre.otto} \chi_t = H(I-Fz)^{-1} G w_t=K(z)w_t.\end{equation}
%Under our assumptions, the innovations and the transfer function  are unique up to  premultiplication and postmultiplication  by orthogonal matrices respectively, so that we can assume with no loss of generality  that $w_t=v_t$  and $K(z)=k(z)$, where $v_t$ and $k(z)$ are defined in \eqref{eq:tre.uno} and \eqref{eq:tre.tre} respectively.

When  $N>N_0\geq q$, so that  $(\chi_t)$ is a singular stochastic process, { i.e. a process with a singular 
spectral density, }
the left inverse of $k(z) $ is not unique. To see this, consider the  Smith-McMillan form (see e.g. \cite{deistler12})
\begin{equation}\label{eq:tre.dieci} k(z) = u(z) d(z) v(z) ,\end{equation}
where $u(z)$ and $v(z) $ are $N\times N$ and $q\times q$, respectively, unimodular polynomial matrices (i.e. their determinants  are   non-zero constants) and 
$$  d(z) = \begin{pmatrix} \frac{\d\epsilon_1(z)}{\d\psi_1(z)} & 0&  \cdots& 0 \\ && \ddots&\\
0& 0 & \cdots & \frac{\d\epsilon_q(z)}{\d\psi_q(z)} \\  0 & 0 & \cdots & 0\\ \vdots  \\  0 & 0 & \cdots & 0\\
\end{pmatrix},$$
where the matrix of zeros at the bottom is $(N-q)\times q$, $\epsilon_i$ and $\psi_i$, for $i=1,\ldots,q$,  are { relatively prime}
monic polynomials, $\epsilon _i$ divides $\epsilon _{i+1}$ and { $\psi_{i+1}$ divides $\psi_{i}$. } Then a particular 
causal left inverse is given by
\begin{equation} \label{eq:tre.undici} h^{-}(z) = v^{-1} (z) (d'(z) d(z)) ^{-1} d'(z) u^{-1} (z).\end{equation}
As is easily seen, $h^{-} (z) $ has no poles or zeros for $|z|\leq 1$ so that
\begin{equation}\label{eq:tre.dodici} v_t = h^- (z)\chi_t\end{equation}
is a causal relation and thus $(v_t) $ are indeed innovations.

In order to uniquely determine $k(z)$ corresponding to the Wold decomposition from $f_\chi$ (compare \eqref{eq:tre.cinque}), in addition  to stability and the miniphase assumption, we have to remove its non-uniqueness caused by post-multiplying by a $q\times q$ orthogonal matrix. This leads to the following assumption guaranteeing uniqueness of $k(z)$:

\begin{ass}\label{BBB} For $q=r$, we assume that $k(0)=I_N$, i.e. $Q_0=I_N$ holds. For $q<r$, we assume that the top $q\times q$ submatrix of $k(0) $ is (1)~non-singular (which is the case generically) and  (2)~lower triangular (with non-zero diagonal elements).
\end{ass}

\subsection{Static and dynamic factors}\label{subsec:tre.due}

 \noindent {    In Section \ref{subsec:tre.uno} we have argued  that
by Assumption \ref{ass:tre}, under a suitable choice of $N_0$, for $N\geq N_0$ the dimension of the space  $\mathcal S^N_t$ spanned by $\chi_{it}$, $i=1,\ldots,N$ is $r$. 
Let 
$ (f_t)$  be an $r$-dimensional  process such that 

\noindent (i) 
$f_t$ 
 forms a basis in  $\mathcal S^N_t$,
 
 \noindent (ii) $f_t = S \chi_t$, with $S$ independent of $t$.
 
 \noindent Of course (i) and  (ii) imply that $(f_t)$ is weakly stationary with a rational spectral density and that
\begin{equation} \label{eq:tre.tredici} \chi_t = L f_t,\end{equation}
where $L$ is an $ N\times r$ matrix independent of $t$.  The  vector $f_t$ is called a vector  of
{\it  (minimal) static factors} and $L $ the corresponding loading matrix.

  As is easy to see,  a minimal static factor is unique  up to premultiplication by a constant non-singular matrix $\cal T$ and the  factor loading matrix $L$ is  unique up to postmultiplication by~${\cal T}^{-1}$.  By \eqref{eq:tre.sette}, the state $x_t$ is a static factor, though in general  not a minimal one.

In particular, let $(f_t)$ be a process of static factors corresponding to $N_0$. Because ${\rm dim} \, \mathcal S^N_t= {\rm dim}\, S^{N_0}_t$ for all $N>N_0$,   $f_t$ is a basis in  $\mathcal S^N_t$ for all $N>N_0$. Thus there exists  a representation \eqref{eq:tre.tredici}  in which the factors are independent of $N$ and the 
 matrix $L$ corresponding to $N$ is nested in the matrix $L$ corresponding to $M$, for $M>N$.  In other words, there 
 exist an $r$-dimensional vector of factors $f_t$   and  a matrix $L\in\mathbb R^{\infty\times r}$ such that 
 $\chi_{it} = L_i f_t,$ for all $i\in \mathbb N$, where $L_i$ is the $i$-th row of $L$.

   Of course, for estimation uniqueness of $f_t$ and $L$ is desirable. A common normalization is to assume that the top $r\times r$ submatrix of $L$, $L_1$ say,   is nonsingular (which is the case, generically)  and then to impose $L_1=I_r$. Clearly this corresponds to the selection of the first $r$ elements of $\chi_t$ as  (minimal) static factors.
   In this case the on- and above-diagonal entries of ${\rm E} f_t f'_t$
  are additional free parameters.   We will  refer to this normalization as the {\it standard normalization.}  Another common normalization is to assume that 
 ${\rm E} f_t f'_t=I_r$. Then $L$ is unique up to multiplication by an orthogonal matrix, which is  made unique by assuming appropriate $Q-R$
decompositions  for $L_1$. 

 A special explicit form of the  static factors can be obtained in the following way. Let $R$  be an $N\times r$ matrix such that $RR' =  \gamma_\chi(0)$, the expected value of $\chi_t{\chi}'_{ t}$. As $\gamma_\chi(0)$  has rank $r$ for $N\geq N_0$, we have ${\rm rank} \, R =r$ for $N\geq N_0$ as well. Now define
\begin{align}
    f_t & = (R'R)^{-1} R' \chi_t\label{eq:nuovissima}\\ &= (R'R)^{-1} R' k(z) v_t= w(z) v_t, \label{eq:chilosa}
    \end{align}
(see \eqref{eq:tre.tre} for the  first  equality in \eqref{eq:chilosa}).  Note that $f_t$ is orthonormal and that   $L=R$.

When $R= P$, where   $P $ has the first~$r$ normalized eigenvectors  of $ \gamma_\chi(0)$ on the columns, the factors are the  first $r$  principal components of $\chi_t$.   Note that if we take, for each $N$,  
$f_t$ as the first $r$   principal components of $\chi^N_t$, the factors depend on $N$ and 
the matrices $L$ are not nested. The same occurs if the principal components are normalized, that is if $R= \sqrt{\Lambda^{-1}} P,$  where $\Lambda $ is the diagonal matrix with the first $r$ eigenvalues of $ \gamma_\chi(0)$ on the diagonal.

Obviously, if the rank of the 
$r\times r$ spectral density of $(f_t)$ were less than $q$, then by \eqref{eq:tre.tredici} Assumption \ref{ass:due} would be violated, so that $r\geq q$ must hold. The spectral density of $(f_t)$ is then 
 nonsingular  or singular, for~$r=q$ or $r>q$ respectively. Thus the dynamics  of  the static factor process $(f_t)$  can be represented by a nonsingular or singular ARMA process of type  \eqref{eq:tre.uno}, respectively.   By \eqref{eq:tre.tredici} 
 and \eqref{eq:chilosa}, the Hilbert spaces spanned by $\chi_t$ and $f_t$ are the same and therefore,  $v_t$ is an innovation for $f_t$ as well
 and $k(z) =Lw(z)$ is the corresponding transfer function.}

  The innovations $(v_t)$ in \eqref{eq:tre.tre}, and \eqref{eq:chilosa}, are called the {\it (minimal) dynamic factors}. As has been stated already, under our assumptions, for given $(f_t)$ they are unique up to  premultiplication  by a non-singular matrix or by an orthogonal matrix if we assume orthonormality of $v_t$.  

Piecing together what we have  seen here and
in the previous section, under Assumption \ref{ass:tre} the latent variables  can be represented as follows:
\begin{align} \chi_{it} & = L_i f_t \label{eq:tre.sedici} \\ \alpha(z) f_t&= \beta(z) v_t,\label{eq:tre.diciassette} \end{align}
for $i\in \mathbb N$, where $L_i$ is the $i$-th row of the matrix $L\in \mathbb R^{\infty\times r}$ and 
$$ \alpha(z) = I_r-\sum _{j=1}^p A_j z^j,\ \  A_j\in \mathbb R^{r\times r},\ \ \  \ \beta(z)=\sum _{j=0} ^m B_j z^j,\ \ B_j\in \mathbb R^{r\times q} , $$
and where we assume that $\alpha(z)$ satisfies the stability condition, $\beta(z)$ satisfies the  
the miniphase condition (that is, respectively,   \eqref{eq:tre.due} with $P(z)$ replaced by $\alpha(z)$ and  \eqref{eq:tre.aggiu} with $Q(z)$ replaced by $\beta(z)$)
and that $\alpha(z)$ and $\beta(z) $ are left coprime as $P(z)$ and $Q(z)$ in Assumption  \ref{ass:febbraio}.

 Lastly, let us point out that  normally  the literature on DFM's assumes, rather than   our Assumption \ref{ass:tre}, 
the existence of representation \eqref{eq:tre.sedici}--\eqref{eq:tre.diciassette}, that is the existence of $N$-independent static factors, see the seminal papers \cite{stockwatson02JASA, stockwatson02JBES}, \cite{baing02}, \cite{bai03}.

 As is seen below,    Assumption \ref{ass:tre}  and  representation \eqref{eq:tre.sedici}--\eqref{eq:tre.diciassette}  are  equivalent.
That  Assumption \ref{ass:tre} implies  representation \eqref{eq:tre.sedici}--\eqref{eq:tre.diciassette} has been proved above.
Conversely,  if $(f_t)$ fulfills equation \eqref{eq:tre.diciassette}, then it has the minimal  state-space representation
$$ \begin{aligned} \tilde x_{t+1} & = \tilde F \tilde x_t + \tilde G v_t\\ f_t &= \tilde H \tilde x_t,\\ \end{aligned}$$
where $\tilde x_t$ is a $\tilde n$-dimensional state  vector.   Thus  the $N$-dimensional vector $\chi_t$ has the 
minimal state-space representation 
$$ \begin{aligned} \tilde x_{t+1} & = \tilde F \tilde x_t + G v_t\\ \chi_t &= L\tilde H \tilde x_t,\\ \end{aligned}$$
where $\tilde n$ is independent of $N$,
so that representation \eqref{eq:tre.sedici}--\eqref{eq:tre.diciassette} implies Assumption \ref{ass:tre}.

Another assumption on the static factors  is usually imposed, i.e.  that the first  $r$ eigenvalues of  the covariance matrix  $\gamma_\chi(0)$ diverge as $N\to \infty$. 
Its  introduction and motivation are better discussed in Section \ref{subsec:PC}, where we deal with estimation of  DFM's.

The advantage of a static factor process $(f_t)$ for $(\chi_t)$ is that  modeling the dynamics of $(\chi_t)$ can be done by modeling the dynamics of $(f_t)$, so that the dimension of the parameter space can be reduced and is independent of $N$. However, even for $q<N$ we may have $r=N$, see Section \ref{subsec:PC}.
% (see e.g. \cite{FHLZ14}).  
This is the case when there is no non-trivial constant  (i.e. independent of $\lambda$) element in the left kernel 
of $f_{\chi} (\lambda)$, as this is equivalent 
to 
$$\gamma_{\chi}(0) =\int_{-\pi}^\pi f_{\chi}(\lambda) d\lambda$$
being nonsingular (for more details see \cite{deistler2019}).

\subsection{Singular ARMA systems: The genericity of the AR case}\label{sec:tre.tre}
In this section  we explain that for the case $r>q$ ``generically'', in a sense to be described below,  the static factor 
process $(f_t)$  in    \eqref{eq:tre.diciassette} is an AR process. This is important because  estimation in the AR case is much simpler compared to the ARMA case. 

Consider the ARMA system \eqref{eq:tre.diciassette}
$$ \alpha (z) f_t=\beta (z) v_t,$$ where $r>q$ and thus $\beta(z)$ is a ``tall'' matrix. As is intuitively clear, $\beta(z)$ is  generically  zeroless (i.e. in the Smith-McMillan form \eqref{eq:tre.dieci} the $\epsilon_i$ are generically constant), since e.g. the zeros of a suitably chosen nonsingular $q\times q$ submatrix of $\beta(z)$ ``typically'' can be compensated  by 
another $q\times q$ submatrix of $\beta(z)$. In other words, and to be more precise, for given $p$ and $m$ (the orders of the AR and MA matrix polynomials respectively), generically, i.e. for an open and dense subset in the parameter space, $\beta(z)$ has no zeros.
 This implies  that generically the ARMA system \eqref{eq:tre.diciassette} fulfills the minimum phase condition or that, equivalently, $v_t$ is an innovation process for $f_t$.

As is well known, every zeroless $r\times q$  polynomial matrix, with $q<r$,   can be extended  to a unimodular 
$r\times r$  matrix  $\tilde \beta(z)=(\beta(z) \ \delta(z))$, say. Now write  \eqref{eq:tre.diciassette}
as
\begin{equation}\label{eq:tre.ventidue}  \tilde \beta^{-1} (z) \alpha (z) f_t =\begin{pmatrix} v_t\\ 0\end{pmatrix} .\end{equation}
Since $\tilde\beta(z)$ is unimodular, $\tilde \beta ^{-1}(z)$ is unimodular too and thus \eqref{eq:tre.ventidue} is a (singular) autoregression, additionally satisfying   the stability condition \eqref{eq:tre.due}.  Multiplying both sides of \eqref{eq:tre.ventidue} by $(\beta (0)\ \delta(0))^{-1}$, which is of course nonsingular, we obtain:
\begin{equation}\label{eq:tre.ventitre} a(z) f_t = b v_t,\ \   a(z)= I_r- \sum_{j=0}^{ \tilde p} A_j z^j,\ \   b\in \mathbb R^{r\times q},\end{equation}
where  $ b=\beta(0)$, see e.g. \cite{andersondeistler08}, \cite{anderson2016}. Thus, in a certain sense, for $r>q$, ``almost every'' factor process $(f_t)$ can be assumed to be generated by a singular  AR process.

As opposed to regular AR processes (where   $r=q$ and $ b =I$ hold), here the assumption  $ a(0)=I$ does not guarantee identifiability, unless the assumptions that $( a(z) \  b)$ is left coprime and that ${\rm rank}\, (A_{\tilde p},  b)=r$ are imposed.

In this AR setting, the factor process is described by the integer-valued parameters $r$,  $q$  and $\tilde p$ (the latter 
being the autoregression order) and the parameter space guaranteeing identifiability is 
\begin{equation}\label{montanario}\begin{aligned} \Theta& = \left \{ {\rm vec} (A_1, \ldots, A_{\tilde p},  b),\  {\rm where}\  \eqref{eq:tre.due}\ {\rm holds},  A_{\tilde p}\ {\rm and}\  b\ {\rm are\ left\ coprime}\right.  \\ &\hskip.6cm \left.  \  {\rm and} \ {\rm rank}(A_{\tilde p},\  b)=r\ \right \}.\end{aligned}\end{equation}
However, not every singular AR system can be described in such a parameter space. For more general parameter spaces see \cite{DEISTLER2011},
 where it has been shown that by prescribing the column degrees of $a(z)$ and by assuming ${\rm rank } \, (A_c, b)=r$, where $A_c$ denotes the column  end matrix,  every singular AR system can be parameterized (see also Section \ref{tantopecanta}).  

It is important to point out that $r>q$  has been invariably observed in empirical applications of DFM's to large macroeconomic datasets, see \cite{BLL21} for a review. 
A very interesting consequence of singularity is briefly accounted for in Section \ref{subsec:quattro.uno}.

%\subsection{Consequences of singularity for Macroeconomic Applications of Factor Models}\label{sec:quattro}

%\noindent This section is Marco's care. 

%\noindent a. This is a Review of Macroeconomic  results obtained by applying 
%Anderson-Deistler  to macroeconomic models.

%\noindent b. I am attaching a  recent paper I coauthored. I would like to 
%make reference to it in this section. See in particular Section 3.2 and Appendix A. 

\section{ Separation of the Common Components. Estimation}\label{Denoising}
\subsection{Principal components}\label{subsec:PC}

We start by illustrating  estimation of the latent variables $\chi_ {it}$, given the observables $y _{it}$,  by means of this simplest  example. Assume that 
\begin{equation}\label{eq:example} y_{it} = \chi_{it} + \xi_{it}, \ \ \ \ \chi_{it} = L_i v_t,\end{equation}
where $(v_t)$  and $(\xi_{it}) $ are scalar unit-variance white noise processes,  fulfilling \eqref{eq:due.due}  and \eqref{eq:due.tre}. Moreover, assume that the processes $(\xi_{it})$ are mutually orthogonal at all leads and lags.  We have:
$$ \begin{aligned} \gamma_\chi(0) &= {\rm E} \chi_t {\chi}'_{ t}= \begin{pmatrix} L_1 & L_2 & \cdots & L_N\end{pmatrix} '   \begin{pmatrix} L_1 & L_2 & \cdots & L_N\end{pmatrix}\\  \gamma_y (0) &={\rm E}\, y_t{y}'_{ t}=
 \gamma_\chi(0) + I_N 
\end{aligned}$$
The model is static and  $f_{\chi}(\lambda) = (2\pi )^{-1}  \gamma_\chi(0)$ for all $\lambda$'s.  All assumptions \ref{ass:uno} through \ref{ass:quattro} and  \ref{ass:febbraio}  are obviously  fulfilled.   Regarding  Assumption \ref{ass:cinque}, 
the rank of $f_{\chi}(\lambda)$ is 1 so that $q= 1$. We have
$$\omega^N_{\chi,1}= (2\pi) ^{-1}\sum _{i=1} ^N L_i ^2, $$
(which is independent of $\lambda$) so that assuming that  $\sum  L_i^2\to \infty$, Assumption \ref{ass:cinque} is also fulfilled. 
Lastly, observe that  the first eigenvalue of the matrix  $\gamma_y(0)$ is
$$ \mu ^N _{y,1} =(2\pi) \omega^N_{\chi,1} +1 =\sum _{i=1}^N L_i^2 +1, $$
with left eigenvector $(L_1\ L_2\ \cdots\ L_N)$.
Now consider the following  average of the observable variables $y_{it}$, $i=1,\cdots,N$:
$$p^N_{y,t} = \frac{1}{\sqrt{\mu^N_{ y,1}}} \sum _{i=1} ^N L_i y_{it},$$
which is known as the first principal component of the $N$-dimensional vector   with coordinates 
$y_{it}$, $i=1,\ldots,N$, and define 
\begin{equation}\label{PC} A^N_{y,t}= \frac{1}{\sqrt{\mu^N_{ y,1}}} p^N_{y,t}.\end{equation}
 We see that 
$$\begin{aligned} A^N_{y,t}&=\frac{1}{{\mu^N_{ y,1}}}  \sum _{i=1} ^NL_i \chi_{it} + 
 \frac{1}{{\mu^N_{ y,1}}}  \sum _{i=1} ^N L_i\xi_{it} = \frac{1}{{\mu^N_{ y,1}}}  \sum _{i=1} ^N L_i^2 v_t + 
 \frac{1}{{\mu^N_{ y,1}}}  \sum _{i=1} ^N L_i \xi_{it}\\
   &=\frac{\sum_{i=1} L_i^2}{\sum _{i=1}^NL_i^2+1} v_t +   \frac{1}{\sum _{i=1}^N L_i^2+1}  \sum _{i=1} ^N L_i \xi_{it},\end{aligned}$$
which implies that 
$${\rm E}  (A^N_{y,t}-v_t)^2= \left [ \frac{\sum_{i=1} L_i^2}{\sum _{i=1}^NL_i^2+1}-1\right ]^2 +  \frac{\sum _{i=1}^N L_i^2}{\left [\sum _{i=1}^N L_i^2+1\right ]^2},$$
and therefore  that the limit in mean square of $ A^N_{y,t}$, as $N\to \infty$, is $v_t$.  Also, in mean square, the projection of 
$y_{it}$ on $A^N_{y,t}$   converges to $\chi_{it}$ and the  regression  coefficient  to $L_i$.

  Example \eqref{eq:example} illustrates 
 the basic features of the estimation techniques used in DFM's. 
The weighted  average  $A^N_{y,t}$,  which is the  rescaled  first principal component of the observable variables $y_{it}$,   does both the ``cleaning'' of the  $y$'s, in that 
it averages out the  idiosyncratic components, and the {\it consistent  estimation}  of 
the common components.

However,  the estimate $A^N_{y,t}$  is  defined using the population  covariances of the $y$'s---as though, so to speak,  $T$ were infinite---and is therefore unfeasible.  In empirical situations, such covariances are estimated  and  depend both on $N$ and $T$.  Moreover, the principal-component technique  must be extended to the  general case, in which, firstly,   $r$, the number of static factors, can be greater than $q$  and $q$ can be greater than unity and, secondly, the static factors  depend on the dynamic factors through an ARMA model.   Before we go over the estimation procedure in the general case we must however introduce another assumption.

\begin{ass}\label{ass:sette} The $r$ largest eigenvalues of the covariance matrix  $\gamma_\chi(0)$ diverge as $N\to \infty$.
\end{ass}
We show below, by means of an example, that Assumption \ref{ass:sette} is not a consequence of Assumption \ref{ass:cinque} and that it is necessary 
for consistent estimation of the space spanned by the factors $f_t$.

Note that if the static factors are orthonormal, $\gamma_\chi(0)=LL'.$  In any case, as the covariance matrix of $f_t$ 
is nonsingular, Assumption \ref{ass:sette} is equivalent to asssuming that the first $r$ eigenvalues of $LL'$ diverge as $N\to \infty.$

Under the assumptions \ref{ass:uno} through \ref{ass:sette}, the first  step of the procedure  consists in the estimation of the integers $q$ and $r$. { This is a non-standard problem because, firstly, both $T$ and $N$ tend to infinity, and, secondly, the factors are estimated, not observed. 
\cite{baing02} provide  a class of information criteria allowing to consistently estimate 
$r$.
In the same vein see \cite{hallinliska07}   for the estimation of  $q$.}

The second step estimates the static factors $f_t$.  By \eqref{eq:due.uno} and \eqref{eq:tre.sedici}, setting $L_i=(L_{i1}\ L_{i2}\ \cdots\ L_{ir})$, 
$$ y_{it} = L_i f_t + \xi_{it}= L_{i1} f_{1t}+ L_{i2} f_{2t} + \cdots+ L_{ir} f_{rt}+\xi_{it}.$$
The first $r$  principal components 
of the observable  variables  $y_{it}$, rescaled as in \eqref{PC},  are computed, based on their estimated covariances, and used   to estimate the space spanned by the  factors $f_t$. The common components $\chi_{it}$ and the loadings $L_i$ are  estimated by regressing the $y$'s on the estimated factor space. \cite{bai03} proves, under some additional technical assumptions, that these estimates  
of the common components  converge in probability to their population counterparts, as both $N$ and $T$ tend to infinity,  with rate
$$\max\left ( \frac{1}{\sqrt{N}},\frac{1}{\sqrt{T}}\right ).$$

In the third step an ARMA model  for the estimated static factors is estimated, see equation \eqref{eq:tre.diciassette}, this leading to the estimation of the dynamic factors as the innovations of the ARMA model.  When $r>q$  the ARMA can be replaced by a singular $AR$, see Section \ref{sec:tre.tre}.  For this approach  see \cite{FGLR09}.

Some observations are in order. 

\noindent (I) Firstly, let us show by an example that Assumption 	\ref{ass:sette}, which is the static counterpart of Assumption \ref{ass:cinque}, is necessary for  consistent estimation by means of principal components.   Let us slightly modify model \eqref{eq:example} in the following way:
\begin{equation} \begin{aligned}\label{eq:sorribes}y_{1t} & =  v_t +v_{t-1} + \xi_{1t} \\ y_{it} &= v_t  + \xi_{it},\ \ {\rm for } \ \ i>1.\end{aligned}\end{equation}
The space  ${\cal S}^N_t$, spanned by $\chi_{it}$, $i\leq N$, for $N>1$, has dimension $2$, so that $r=2$, a basis  being $f_{1t}=v_t$, $f_{2t}= v_{t-1}$.
It is fairly  easy to see that  both the first eigenvalue of $\gamma_\chi(0) $ and of $f_{\chi}(\lambda)$  diverge  at rate $N$.   However, the second eigenvalue of $\gamma_\chi(0) $ is bounded. Thus Assumption \ref{ass:cinque} holds with $q=1$ but Assumption \ref{ass:sette} does not. It also easy to see that 
the  first   rescaled principal component
converges to $v_t$, but 
the second  does not   ``clean''  the variables $y_{it}$ from the idiosyncratic component and therefore does not converge  to the space spanned by the factors.  The consequence is that  the  common and idiosyncratic   components of $y_{it}$, as estimated by principal components,   are, respectively:
$$\begin{aligned}& \hbox{
$v_t$  and $v_{t-1}+\xi_{1t}$ for $i=1$,}\\
&\hbox{
$v_t$  and $\xi_{1t}$ for $i>1$.}\\ \end{aligned}$$
For another example 
consider 
\begin{equation}\label{eq:ventinove}\chi_{it} = v_t + M_i v_{t-1}\end{equation}
with $\sum _{i=1} ^\infty M_i^2 < \infty$.  In this case the common  and idiosyncratic components of $y_{it}$ estimated by  the principal components are $v_t$ and $M_iv_{t-1}+ \xi_{it}$, respectively.

%\noindent { (II) Consider again example \eqref{eq:sorribes}.  Assumptions 
%\ref{ass:uno}, \ref{ass:due}, \ref{ass:quattro} and  \ref{ass:cinque}, plus 
%\eqref{eq:due.due} and \eqref{eq:due.tre}, are all fulfilled, so that, by the result in  \cite{fornilippi01} mentioned in Section \ref{sec:due}, the decomposition in  \eqref{eq:sorribes} is unique. On the other hand, as we have seen above, the principal component  technique would 
%obtain $v_t$  as the latent variable for all $i$ and the noise
% v _{t-1}+u_{1t}\ u_{2t}\ u_{3t}\ \cdots,$$
%thus the same as in \eqref{eq:sorribes} for $i>1$ but $v_{t-1}+u_{1t}$ %for $i=1$.
%Of course the first eigenvalue of this  alternative noise is bounded so that Assumptions \ref{ass:uno}, \ref{ass:due}, \ref{ass:quattro} and  \ref{ass:cinque} are fulfilled. However, there is no contradiction with the uniqueness result because \eqref{eq:due.tre}, i.e. orthogonality of 
%the noise terms with the latent variables {\it at any lead and lag}, does not hold. 
%}

\noindent{(II)  What examples \eqref{eq:sorribes} and 
\eqref{eq:ventinove} show is that in order to ``finally'' (i.e. for $N\to \infty$) separate the idiosyncratic from the common components it is necessary that  each factor is loaded infinitely often  and that the loading coefficients, if declining, do not decline too fast. This is usually rendered by  saying that the factors must be ``pervasive''  and has a precise formulation in 
Assumption \ref{ass:sette}.}

%\noindent (III) Lastly,  let us remark that what converges  in the estimation procedure outlined above  are the estimated common components and, as an obvious consequence, the estimated idiosyncratic components.  
%%converge to the space  spanned by $f_t$. 

%\subsection{Estimation in the frequency domain}\label{FDE}
%In a certain sense, Assumption \ref{ass:tre} says that the dynamics of %the latent processes are bounded in  $N$, rather than increasing with %$N$. 

\subsection{Generalized Dynamic Factor Models}\label{appiccicata}

 Let us conclude this section by mentioning  a strand of literature in DFMs in which Assumption \ref{ass:tre} does not necessarily hold.  As we have seen, Assumption \ref{ass:tre} is equivalent to assuming  that 
the dynamics of the latent variables $\chi_{it}$ are completely accounted for by the dynamics of the finite-dimensional, $N$-independent vector  $f_t$, {\it via} the static loadings $L_i$. 
The following factor model is an elementary example in which no static factors exist.
Let 
\begin{equation}\label{eq:dinamico} \chi_{it} = \frac{1}{1-\alpha_iz} v_t ,\end{equation}
where $v_t$ is scalar unit-variance  white noise, $-0.8\leq \alpha_i\leq 0.8$, $\alpha_i\neq \alpha _j$ for all $i$ and $j$, $i\neq j$.
We have  $f_{\chi}(\lambda) = (2\pi)^{-1}  H_N(\lambda) H^*_N(\lambda)$, where
$$ H_N(\lambda)= 
\begin{pmatrix} (1-\alpha_1e^{-i\lambda})^{-1} & \cdots &  (1-\alpha_Ne^{-i\lambda})^{-1}\end{pmatrix}'.$$
 As $f_{\chi}(\lambda)$ has rank one for all $\lambda$, the first eigenvalue of  $f_{\chi}(\lambda)$ is its trace, that is 
$$  \omega^N_{\chi,1}(\lambda)= (2\pi)^{-1}\sum _{j=1}^N |1-\alpha_ie^{-i\lambda}|^{-2}.$$
Because  $\omega ^N_{\chi,1}(\lambda)$ diverges for all $\lambda$,  we have $q=1$.  On  the other hand, 
$$\chi _{it} = v_t + \alpha_i v_{t-1} +  \cdots +\alpha _i^{N-1} v_{t-N+1} + \cdots.$$
If the matrix  $ \gamma_\chi (0)={\rm E} \chi_t \chi_t' $   were singular then
the $N\times N$  matrix with $\alpha_i ^{m-1}$ in entry $(i,\ m)$, with $i,m=1,\ldots N$  should be singular.  But the determinant 
of the latter  is the Vandermonde determinant  of $\alpha_1,\ \ldots, \ \alpha_{N-1}$, which vanishes only if  at least 
two of the $\alpha $'s are equal. Thus the dimension  of the space ${\cal S}^N_t$  is $N$, not  some  $N$-independent  $r$, and Assumption \ref{ass:tre} does not hold.  As a consequence the estimation technique based on   a fixed finite number of principal components does not apply.

{ The DFM without Assumption \ref{ass:tre}  has been  studied by means of frequency-domain methods in \cite{forni_generalized_2000},  \cite{fornilippi01}, \cite{HALLIPPI13}, \cite{FHLZ14}, \cite{FHLZ15}, \cite{FoGiLiSo}, and called Generalized Dynamic Factor Model. The main tool is the dynamic principal component analysis introduced in \cite{BRILL81}, which consist of linear combinations of current, past and future values 
 of the observable variables $y_{it}$ (instead of just current values as in the standard principal components). 
 
 We cannot discuss here the merits of this ``dynamic'' approach relative to the one adopted in the present paper. We limit ourselves to observing that by means of the   dynamic principal components the latent variables in model \eqref{eq:dinamico}   can be consistently estimated. Moreover,  by means of the dynamic principal components, 
 the common and idiosyncratic  components of the variable $y_{1t}$  in example \eqref{eq:sorribes} would be correctly estimated as~$v_t+v_{t-1}$ and $\xi_{1t}$ respectively. The same holds for example \eqref{eq:ventinove},  where  by means of the dynamic principal components we estimate the latent variables $v_t+M_i v_{t-1}$.} 
 Thus the  approach based on the dynamic principal components gives the correct results even  when Assumption \ref{ass:tre} does not hold, or  when   Assumption \ref{ass:tre}  holds but not Assumption
 \ref{ass:sette}.  
 
% Lastly, let us point out that the terminology used in the literature on  dynamic factor models is far from being  
% uniform across different papers.
%An example is: ``Generalized Dynamic Factor Model''  used when Assumption \ref{ass:tre} holds,  the ``generalization''  (with respect to exact models)
%    being the possibility of cross-correlated idiosyncratic terms, or ``static model'' used instead of ``model in which the common components have a static representation in the factors''.  
%     In some cases the terminology is even misleading, an outstanding example being 
%     ``static factors''   and ``dynamic factors'', which are  used to denote, respectively,  a vector process with 
%    an ARMA structure, see \eqref{eq:tre.diciassette},   and a white-noise vector process. 
%   In the present paper we only consider models fulfilling Assumptions \ref{ass:uno} through \ref{ass:sette}, Assumption \ref{ass:tre} in particular, and call them Large-Dimensional Dynamic Factor Models.
% 
 
\subsection{ A State-Space Formulation of a DFM. Generic Identifiability and Maximum Likelihood Estimation
}\label{sec:quattro.due}

\subsubsection{The State-Space Formulation}
A different approach to estimation of DFM's has been introduced in 
\cite{DGRqml}. The paper employs a maximum likelihood estimator for the DFM resulting from the assumption that the idiosyncratic components are  cross-sectionally uncorrelated, and shows that  this misspecification has no effect on the estimated common components as $N\to \infty$. See also  \cite{Bai_Li2016}, \cite{Barigozzi_Luciani2019} and  \cite{PONCELA2021}.
This motivates the following formulation of a DFM in state space.

%In this section a DFM in state-space formulation is described . 

To repeat, it is assumed that the underlying model is  an exact factor model, i.e. that the univariate idiosyncratic components  are mutually uncorrelated; in addition we assume that they are of AR(1)  type (the latter  can easily be generalized). Both assumptions of course restrict generality, but are nevertheless  appropriate for many applications. An advantage of the state-space 
formulation  is that an EM algorithm of Shumway-Stoffer type, see \cite{shumway2000},  can be used for parameter estimation  by means of the Kalman smoother. Clearly in this case identifiability  is an important advantage.

We retain the assumption that $N>r\geq q$ and as earlier, we have the latent variables and minimal static factors related by \eqref{eq:tre.tredici}. Further, in case $r>q$, and relying on an assumption of genericity, there is no loss of generality in working with an AR model for the minimal static factor process as given by \eqref{eq:tre.ventitre}. Indeed, even if $r=q$, we shall assume that such a model is valid. This, of course, is not a  consequence of genericity, and is restrictive. 

Next, we shall assume that the $i$-th entry of the idiosyncratic component, $\xi_{it}$, is the first order AR process:

\begin{equation}\label{eq:ui}
\xi_{i,t}=\delta_i\xi_{it-1}+\eta_{it}
\end{equation}
where $|\delta_i|<1, \quad i=1,2,\dots,N$, and $(\eta_{it})$ are mutually uncorrelated zero-mean white noise processes, and also uncorrelated with the process $v_t$ driving the factor process model.

These assumptions follow the construction of a state-space model, where the $(r\tilde p+N)$-dimensional state vector is taken to be 
\begin{equation}\label{eq:nuovissima1}
 x_t=\begin{bmatrix}
    f_t\\f_{t-1}\\\vdots\\f_{t-\tilde p+1}\\ \xi_t
    \end{bmatrix}
\end{equation}
The model is given by
\begin{equation}\label{eq:SVmodel}\begin{aligned}
x_{t+1}=&\begin{bmatrix}
A_1&A_2&\dots&A_{\tilde p-1}&A_{\tilde p}&0\\
I&0&\dots&0&0&0\\0&I&\dots&0&0&0\\
\vdots&&&&&\vdots\\0&0&\dots&I&0&0\\
0&0&\dots&0&0&\delta\end{bmatrix}x_t+\begin{pmatrix} b& 0 \\ 0& 0\\ 0& 0 \\ \vdots\\ 0 & 0 \\ 0&I_N \end{pmatrix}\begin{pmatrix}v_{t+1} \\ \eta_t\end{pmatrix}\\
&=Ax_t+B\begin{pmatrix}v_{t+1} \\ \eta_t\end{pmatrix} \\
y_t&=[L\;\;0\;\;I_N]x_t=Cx_t
\end{aligned}\end{equation}
Here, 
\[
\delta={\rm{diag}}(\delta_1,\delta_2,\dots,\delta_N),\ \ \ \eta_t= (\eta_{1t}\ \eta_{2t}\ \cdots \ \eta_{Nt})',\ \ \ \xi_t= (\xi_{1t}\ \xi_{2t}\ \cdots \ \xi_{Nt})'
\]
and, to repeat, 
%\begin{equation}\label{dragoni}
\begin{equation}\label{AAA}a(z)f_{t+1} =bv_{t+1},\ \ \ a(z)=I_r-\sum_{j=1}^{\tilde p} A_jz^j.\end{equation}
Of course, we retain the stability requirement that  ${\rm{det}}(I_r-\sum_jA_jz^j)\neq 0$ for $z\leq 1$. 

Note that the dimension of the state depends on $N$ and  that   this may cause problems in proving consistency, see 
\cite{Banbura2014}.

\subsubsection{Generic Identifiability}\label{tantopecanta}
In studying  identifiability of such a model, one should eliminate unnecessary parameters. Hence we shall assume, as in Section \ref{subsec:tre.due},  the standard normalization for  $L$   
(the top $r\times r$ submatrix of $L$ is  equal to $I_r$),  in order to uniquely obtain $L$  from  $ {\rm E} \chi_t\chi'_t$. 

We shall  assume, using a further appeal to genericity, that none of the quantities $\delta_i^{-1}$ is a zero of ${\rm{det}}(I_r-\sum_j A_jz^j)$.

The first step in establishing generic identifiability, is to explain how  the {\it separation of common and idiosyncratic components} can be achieved, or,  equivalently, how we can separate the spectrum matrix $f_y$ into its two additive components $f_{\chi}$ and $f_\xi$. There are in fact two ways in which this can be done. 

First,  since the power spectrum $f_y$ is rational, it has a partial fraction expansion. By genericity, each pole is simple. Each $\delta_i$ gives rise to a pole $\delta_i^{-1}$ and appears in the $i$-th diagonal entry of $f_\xi$, but not in $f_{\chi}$, for which all poles are zeros of ${\rm{det}}(I_r-\sum_jA_jz^j)\neq 0$. Hence the residue matrix associated with the pole $\delta^{-1}_i$ in the partial fraction expansion of $f_y$ is a diagonal matrix of rank 1. On the other hand, the residue matrix associated with any pole arising as a zero of ${\rm{det}}(I_r-\sum_jA_jz^j)\neq 0$ will generically be a matrix with many, and maybe all, nonzero entries, even should it have rank 1. Hence the power spectrum $f_{\chi}$ can be determined by adding together those summands of the partial fraction expansion of $f_y$ whose residue matrices are other than diagonal and of rank 1.

For the alternative procedure, let us suppose that $N$ exceeds $2q$. We can expect by genericity that $q\times q$ submatrices of $f_{\chi} $ obtained by deleting an arbitrary set of $N-q$ columns and an arbitrary set of $N-q$ rows are nonsingular, while  $(q+1)\times (q+1)$ submatrices obtained via a like process are necessarily singular. 
%{\COLMAN Manfred: Can you give a nice argument for the conclusion that generically \textit{any} $q\times q$ submatrix of $f^N_{\chi}$ is nonsingular? Separately, you thought we used this in some paper. Do you know what the paper was? }
Now for each diagonal entry of $f_y$, choose a $(q+1)\times (q+1)$ matrix of $f_y$ by selecting $(q+1)$ not necessarily continguous rows and $(q+1)$ not necessarily contiguous columns containing that diagonal entry but containing no other diagonal entry. Note that such a choice is possible precisely because $N>2q$. The entries of the submatrix will be identical with the entries of the corresponding submatrix of $f_{\chi}$, save for the entry corresponding to the single diagonal entry of $f^N_y$. Singularity of the submatrix of $f_{\chi}$ for which all but one entry are known will allow identification of the remaining entry, which is a diagonal entry of $f_{\chi}$. Since all diagonal entries of $f_{\chi}$ can be obtained this way, and the off-diagonal entries of the matrix are identical with those of $f_y$, again the separation is achieved.

The next step in establishing generic identifiability is  to construct  the ``real-valued'' parameters $L,\ A_1,\ \ldots,\  A_{\tilde p},\ b,\  \delta$ and ${\rm E}\eta_{it}^2$, for given integral-valued specification parameters $r,\ q,\ \tilde p$ from the given spectral density $f_\chi$, or, equivalently, from the second moments of $(\chi_t)$. This is done as follows:

\noindent 1. From ${\rm E} \chi_t\chi_t'=LL'$, $L$ can be uniquely determined using the  standard normalization introduced in Section  
\ref{subsec:tre.due}.

\noindent 2. The transformation \eqref{eq:nuovissima}  then uniquely defines the second moments of the process $(f_t)$. 

\noindent 3. Now consider the autoregression \eqref{eq:tre.ventitre}; then, as well known, the parameters 
$A_1,\ \ldots,\ A_{\tilde p},\ b$ are uniquely defined from the (population) second moments of $(f_t)$ if the following assumption  holds:

\begin{ass}\label{degregori} 
\begin{equation}\label{dalla1} {\rm E} \begin{pmatrix} f_{t-1}\\ \vdots\\ f_{t-\tilde p}\end{pmatrix}\begin{pmatrix} f_{t-1}\\ \vdots\\ f_{t-\tilde p}\end{pmatrix}'>0\end{equation}
holds.
\end{ass}

\noindent 4. Finally, from \eqref{eq:due.quattro}  we obtain $f_\xi$ from $f_y$ and $f_\chi$ and thus the parameters $\delta $ and ${\rm E}\eta_{it}^2$.

Note that Assumption \ref{degregori} is equivalent to controllability of \eqref{eq:SVmodel}.  Due to our assumptions, \eqref{dalla1} is fulfilled, as easily shown, for $r=q$. For $r>q$, however, which in a certain sense is standard, this may not be the case.  As shown in \cite{DEISTLER2011}, see p. 20,
%
% to tie together state space descriptions of the spectral matrix $f_{\chi}$ and of the transfer function matrix $k(z)$ of the Wold decomposition, i.e., commencing  from $k(z)$, the question of generic identifiability  is whether, generically, the restrictions in $(A,\ B,\  C)$ in \eqref{eq:SVmodel} uniquely determine $(A,\ B,\  C)$ for given $k(z)$. As is well known, if $(A,\ B,\  C)$ is minimal, i.e. controllable and observable, then $(\bar A,\ \bar B,\  \bar C)$ and $(A,\ B,\  C)$  have the same transfer function if and only if there exists a non-singular constant matrix $T$ such that 
%\begin{equation} \label{draghini} \bar A=TAT^{-1}, \ \ \ \bar B= TB,\ \ \ \bar C= CT^{-1}\end{equation}
%hold.
%
%As can be shown, the system  \eqref{eq:nuovissima1}, \eqref{eq:SVmodel} is observable if and only if $A_{\tilde p}$ is non-singular. Now, controllability is equivalent to ${\rm E} x_tx_t'$ being non-singular. Due to our assumptions, this is the case if and only if the one-dimensional components of $(f_t',\ \ldots, f_{t-\tilde p+1}')$  are linearly independent.   For $r=q$, it can be easily shown that this is always the case. For $r>q$, however,  which in a certain sense is standard, this may not be the case.  
%As shown in \cite{DEISTLER2011}, 
in this case a first basis of elements of    $(f_t',\ \ldots, f_{t-\tilde p+1}')$ can be selected and this corresponds to a prescription of column degrees $p_i\leq \tilde p$, $i=1,\ldots,r$, for  $a(z)$ in \eqref{AAA}. With the corresponding prescription of a state vector, this modified state space system is controllable (for this  argument see also the comment on \eqref{montanario} in Section \ref{sec:tre.tre}). 

 \section{Macroeconomic Applications: Some Consequences of Singularity}\label{sec:quattro}{
 {
 A large literature has used DFM's  as a tool for forecasting key macroeconomic indicators, see the seminal papers
\cite{stockwatson02JASA, stockwatson02JBES}, see also \cite{FHLR05} and  \cite{stockwatson16}. In another important application DFM's have been used in structural macroeconomic  analysis. It has been shown that by replacing the macroeconomic variables of interest with their common components, estimated from a large dataset by the DFM technique, provides a solution to a much-debated difficulty known among macroeconomists  as the ``fundamentalness problem''.   Such solution, as we see below, 
depends on the singularity of the static factors and the results 
presented in Section \ref{sec:tre.tre}. 

Interesting issues, arising with nonstationarity of the variables
$y_{it}$, which is of course the case for the majority of the 
macroeconomic variables, are briefly introduced in Section \ref{NS}.
 }
 \subsection{Applications to Structural Macroeconomic  Analysis}\label{subsec:quattro.uno}

 We give here a short illustration of this literature by means of a very simple example. Consider a DFM with $q=1$ and suppose that \begin{equation}\label{eq:cinque.uno}y_{it}=v_t+M_i v_{t-1} + \xi_{it}.\end{equation}
Then focus on 
the vector  $(\chi_{1t} \ \chi_{2t})':$
\begin{equation}\begin{aligned} \chi _{1t}& = v_t+ M_1 v_{t-1}=(1+M_1 z) v_t \\ \chi _{2t} &= v_t+ M_2 v_{t-1}=(1+M_2z)v_t.\\ \end{aligned}\label{eq:singular431}\end{equation}
This is a singular vector and we see that the $2\times 1$ matrix 
$$\begin{pmatrix}1+M_1 z\\ 1+M_2 z\end{pmatrix}$$ is zeroless unless $M_1=M_2$, and thus generically zeroless 
as $(M_1\ M_2)$  varies in an open set of $\mathbb R^2$.  
It is convenient to exclude 
from the parameter space all points $(M_1\ M_2)$ with $|M_1|=1$ or $|M_2|=1$.

Thus generically  the minimum phase condition is fulfilled 
for \eqref{eq:singular431}, or,  in an alternate terminology, $v_t$ is {\it fundamental} in \eqref{eq:singular431}. Note that this does not imply that $|M_1|<1$ or $|M_2|<1$.   In other words, if $M_1\neq M_2$,  $v_t$  is fundamental for the $2$-dimensional vector $(\chi_{1t})$ even though it is non-fundamental   for each of the scalar processes  $\chi_{1t}$ and  $\chi_{2t}$ taken separately. 

 Now suppose that  an econometrician is interested in  $y_{1t}$ and, for the sake of simplicity, that 
 $y_{1t}$ is observed without error, i.e. $y_{1t}=\chi_{1t}$.  We assume also that $y_{1t}=v_t+M_1 v_{t-1}$ is a structural
 equation, i.e. that the parameter $M_1$ and the white noise $v_t$ have a   structural interpretation.
 
 Standard Var analysis would estimate a VAR for 
 $y_{1t}$, which is just an AR in  this case, then the AR would be inverted.  As the generating process is an MA(1), this procedure estimates consistently an MA(1):
 $$ y_{1t} = w _t + N_1 w_{t-1}.$$
 Now, 
 $w_t$, being the residual of an AR,  is an innovation for $y_{1t}$,   which implies that 
 $|N_1|<1$.  Thus
  $N_1$ is equal to $M_1$ only if $|M_1|<1$, otherwise
$N_1=1/ M_1$.  The so-called {\it fundamentalness problem} in Structural VAR analysis  arises because usually the econometrician's information is not sufficient to identify   the structural model among those consistent 
with the spectral density  of the observable vector. In our case the econometrician is not able to decide between 
$N_1$, which is by definition less than unity in modulus, and $1/N_1$.

The solution of the fundamentalness  problem based on DFM's can be presented, in the case of our simple example, as follows:

\noindent 1. We have  assumed that   $y_{1t}$, the variable of interest,  belongs to a large  macroeconomic dataset $(y_{it}),$  $i=1,\ldots , N$.

\noindent 2.   Assuming that the variables in the dataset have the  DFM  structure \ref{eq:cinque.uno},  with $q=1$ and $r=2$,  we apply the separation-estimation technique 
 outlined in Section \ref{Denoising}, thus obtaining the static  factors $f_t$,  the loadings $L_i$ and 
 the common components $\chi_{it}$. 
 
 \noindent 3.   Now consider any $2$-dimensional vector $(\chi_{1t}\ \chi_{it})= ( y_{1t}\ \chi_{it})$, with $i\neq 1$, for example $\chi_t=(y_{1t}\ \chi_{2t})$. An estimate of a singular  VAR 
 for $\chi_t$ and its inversion will consistently estimate a vector MA(1) for $\chi_t$, with a white noise $w_t$ which is fundamental.  On the other hand, $v_t$ is   generically fundamental 
 in \eqref{eq:singular431}. Uniqueness of fundamental representations implies   that  $w_t=v_t$ and the first equation in the estimated vector  MA(1)
 is precisely $y_{1t}= v_t +M_1 v_{t-1}$. 
 
 Note that in step 3 we estimate a VAR for the common components 
 of $\chi_{1t}$ and $\chi_{2t}$.  Alternatively, we  can estimate 
 a singular VAR for the factors.
 For  details in the general case and macroeconomic applications see
 \cite{FGLR09}, \cite{stockwatson16}, \cite{FGLS2020}.

%\section{The Genericity of the AR Case for Singular ARMA Systems}\label{sec:tre}
%\noindent This section is Manfred's care.
%
%\noindent a. This section was  Subsection 3.3 in the your manuscript version. I propose that we have 
%Section 4 on the topic. Basically, here we have (i)~zero-free MA matrix polynomials have a finite left inverse, (ii)~generically, MA matrix polynomials are zero-free. So, this is Anderson and Deistler and coauthors.  
%
%\noindent b. An important corollary is that generically MA matrix polynomials are minimum 
%phase, i.e. fundamental.

\subsection{Nonstationary DFMs and cointegration of the factors}\label{NS}
In general only some of the processes in   an empirical  dataset  are stationary. Assuming that the nonstationary  processes are I(1), the separation-estimation procedure  described in Section \ref{Denoising} applies to the dataset obtained by taking first differences of the I(1) processes. 

Suppose for simplicity that all the processes  $y_{it}$, $\chi_{it}$, $f_t$ and $\xi_{it}$  are I(1).
We consistently  estimate $(1-z)f_t$, $(1-z) \chi_{it}$ and $(1-z) \xi_{it}$. Then the levels are obtained by integration. This, apart from minor issues regarding the initial conditions, is  a 
fairly trivial extension. However, if we want to estimate a VAR for the factors $f_t$ or a vector 
of common components, as in Section  \ref{subsec:quattro.uno}, cointegration 
must be taken into account. Indeed, under the assumption $r>q$, i.e. under singularity of $f_t$, 
the spectral density of $f_t$ has rank $q$ at all frequencies and therefore at frequency zero, so that $f_t$ is cointegrated  with cointegration rank at least $r-q$.

For cointegration of singular vector processes and 
the singular version of the Granger representation theorem, see
\cite{DeistlerWagner17}, \cite{BLL20}. See also  \cite{BLL21} for estimation
and some empirical  applications.}

\bibliographystyle{chicago}

\bibliography{ADL_C}

\newcommand{\noop}[1]{}
\begin{thebibliography}{}

\bibitem[\protect\citeauthoryear{Anderson and Deistler}{Anderson and
  Deistler}{2008}]{andersondeistler08}
Anderson, B. D.~O. and M.~Deistler (2008).
\newblock Generalized linear dynamic factor models--{A} structure theory.
\newblock {\em IEE Conference on Decision and Control\/}.

\bibitem[\protect\citeauthoryear{Anderson, Deistler, Felsenstein, and
  Koelbl}{Anderson et~al.}{2016}]{anderson2016}
Anderson, B. D.~O., M.~Deistler, E.~Felsenstein, and L.~Koelbl (2016).
\newblock {The structure of multivariate AR and ARMA systems: Regular and
  singular systems; the single and the mixed frequency case}.
\newblock {\em Journal of Econometrics\/}~{\em 192\/}(2), 366--373.

\bibitem[\protect\citeauthoryear{Bai}{Bai}{2003}]{bai03}
Bai, J. (2003).
\newblock Inferential theory for factor models of large dimensions.
\newblock {\em Econometrica\/}~{\em 71}, 135--171.

\bibitem[\protect\citeauthoryear{Bai and Li}{Bai and Li}{2016}]{Bai_Li2016}
Bai, J. and K.~Li (2016).
\newblock Maximum likelihood estimation and inference for approximate factor
  models of high dimension.
\newblock {\em The Review of Economics and Statistics\/}~{\em 98\/}(2),
  298--309.

\bibitem[\protect\citeauthoryear{Bai and Ng}{Bai and Ng}{2002}]{baing02}
Bai, J. and S.~Ng (2002).
\newblock Determining the number of factors in approximate factor models.
\newblock {\em Econometrica\/}~{\em 70}, 191--221.

\bibitem[\protect\citeauthoryear{Ba\'nbura and Modugno}{Ba\'nbura and
  Modugno}{2014}]{Banbura2014}
Ba\'nbura, M. and M.~Modugno (2014).
\newblock {Maximum Likelihood Estimation Of Factor Models On Datasets With
  Arbitrary Pattern Of Missing Data}.
\newblock {\em Journal of Applied Econometrics\/}~{\em 29\/}(1), 133--160.

\bibitem[\protect\citeauthoryear{Barigozzi, Lippi, and Luciani}{Barigozzi
  et~al.}{2020}]{BLL20}
Barigozzi, M., M.~Lippi, and M.~Luciani (2020).
\newblock Cointegration and error correction mechanisms for singular stochastic
  vectors.
\newblock {\em Econometrics\/}~{\em 8\/}(1).

\bibitem[\protect\citeauthoryear{Barigozzi, Lippi, and Luciani}{Barigozzi
  et~al.}{2021}]{BLL21}
Barigozzi, M., M.~Lippi, and M.~Luciani (2021).
\newblock Large-dimensional dynamic factor models: Estimation of
  impulse–response functions with {I(1)} cointegrated factors.
\newblock {\em Journal of Econometrics\/}~{\em 221\/}(2), 455--482.

\bibitem[\protect\citeauthoryear{Barigozzi and Luciani}{Barigozzi and
  Luciani}{2019}]{Barigozzi_Luciani2019}
Barigozzi, M. and M.~Luciani (2019).
\newblock Quasi maximum likelihood estimation of non-stationary large
  approximate dynamic factor models.
\newblock Papers, arXiv.org.

\bibitem[\protect\citeauthoryear{Brillinger}{Brillinger}{1981}]{BRILL81}
Brillinger, D.~R. (1981).
\newblock {\em Time Series: Data Analysis and Theory}.
\newblock San Francisco: Holden {Day}.

\bibitem[\protect\citeauthoryear{Burt}{Burt}{1909}]{burt09}
Burt, C.~L. (1909).
\newblock Experimental texts of general intelligence.
\newblock {\em The British Journal of Psychology\/}~{\em 3\/}(1-2), 94--177.

\bibitem[\protect\citeauthoryear{Chamberlain}{Chamberlain}{1983}]{Chamberlain}
Chamberlain, G. (1983).
\newblock {Funds, Factors and Diversification in Arbitrage Pricing Models}.
\newblock {\em Econometrica\/}~{\em 51\/}(5), 1281--1304.

\bibitem[\protect\citeauthoryear{Chamberlain and Rothschild}{Chamberlain and
  Rothschild}{1983}]{chamberlainrotshild83}
Chamberlain, G. and M.~Rothschild (1983).
\newblock Arbitrage, factor structure, and mean-variance analysis on large
  asset markets.
\newblock {\em Econometrica\/}~{\em 51\/}(5), 1281--304.

\bibitem[\protect\citeauthoryear{Deistler}{Deistler}{2019}]{deistler2019}
Deistler, M. (2019).
\newblock Singular {ARMA} systems: {A} structure theory.
\newblock {\em Numerical Algebra, Control and Optimization\/}, 383.

\bibitem[\protect\citeauthoryear{Deistler, Anderson, Filler, Zinner, and
  Chen}{Deistler et~al.}{2010}]{andersondeistlersingular}
Deistler, M., B.~D.~O. Anderson, A.~Filler, C.~Zinner, and W.~Chen (2010).
\newblock Generalized linear dynamic factor models: An approach {\it via}
  singular autoregressions.
\newblock {\em European Journal of Control\/}, 211--224.

\bibitem[\protect\citeauthoryear{Deistler, Filler, and Funovics}{Deistler
  et~al.}{2011}]{DEISTLER2011}
Deistler, M., A.~Filler, and B.~Funovics (2011).
\newblock {AR} systems and {AR} processes: The singular case.
\newblock {\em Communications in Information and Systems\/}~{\em 11}, 225--236.

\bibitem[\protect\citeauthoryear{Deistler and Scherrer}{Deistler and
  Scherrer}{2018}]{deischerrer18}
Deistler, M. and W.~Scherrer (2018).
\newblock {\em Modelle der {Zeitreihenanalyse}}.
\newblock Chem: Birkh{\" a}user {Springer}.

\bibitem[\protect\citeauthoryear{Deistler and Wagner}{Deistler and
  Wagner}{2017}]{DeistlerWagner17}
Deistler, M. and M.~Wagner (2017).
\newblock {Cointegration in singular ARMA models}.
\newblock {\em Economics Letters\/}~{\em 155\/}(C), 39--42.

\bibitem[\protect\citeauthoryear{Doz, Giannone, and Reichlin}{Doz
  et~al.}{2012}]{DGRqml}
Doz, C., D.~Giannone, and L.~Reichlin (2012).
\newblock A quasi maximum likelihood approach for large approximate dynamic
  factor models.
\newblock {\em The Review of Economics and Statistics\/}~{\em 94\/}(4),
  1014--1024.

\bibitem[\protect\citeauthoryear{Forni, Gambetti, Lippi, and Sala}{Forni
  et~al.}{2020}]{FGLS2020}
Forni, M., L.~Gambetti, M.~Lippi, and L.~Sala (2020).
\newblock Common component structural {VAR}s.
\newblock Working {Paper}. {CEPR}.

\bibitem[\protect\citeauthoryear{Forni, Giannone, Lippi, and Reichlin}{Forni
  et~al.}{2009}]{FGLR09}
Forni, M., D.~Giannone, M.~Lippi, and L.~Reichlin (2009).
\newblock Opening the black box: {S}tructural factor models versus structural
  {VAR}s.
\newblock {\em Econometric Theory\/}~{\em 25}, 1319--1347.

\bibitem[\protect\citeauthoryear{Forni, Giovannelli, Lippi, and Soccorsi}{Forni
  et~al.}{2018}]{FoGiLiSo}
Forni, M., A.~Giovannelli, M.~Lippi, and S.~Soccorsi (2018).
\newblock {Dynamic factor model with infinite‐dimensional factor space:
  Forecasting}.
\newblock {\em Journal of Applied Econometrics\/}~{\em 33\/}(5), 625--642.

\bibitem[\protect\citeauthoryear{Forni, Hallin, Lippi, and Reichlin}{Forni
  et~al.}{2000}]{forni_generalized_2000}
Forni, M., M.~Hallin, M.~Lippi, and L.~Reichlin (2000, November).
\newblock The generalized dynamic-factor model: Identification and estimation.
\newblock {\em Review of Economics and Statistics\/}~{\em 82\/}(4), 540--554.

\bibitem[\protect\citeauthoryear{Forni, Hallin, Lippi, and Reichlin}{Forni
  et~al.}{2005}]{FHLR05}
Forni, M., M.~Hallin, M.~Lippi, and L.~Reichlin (2005).
\newblock The {G}eneralized {D}ynamic {F}actor {M}odel: One sided estimation
  and forecasting.
\newblock {\em Journal of the American Statistical Association\/}~{\em 100},
  830--840.

\bibitem[\protect\citeauthoryear{Forni, Hallin, Lippi, and Zaffaroni}{Forni
  et~al.}{2015}]{FHLZ14}
Forni, M., M.~Hallin, M.~Lippi, and P.~Zaffaroni (2015).
\newblock Dynamic factor models with infinite-dimensional factor spaces:
  One-sided representations.
\newblock {\em Journal of Econometrics\/}~{\em 185}, 359--371.

\bibitem[\protect\citeauthoryear{Forni, Hallin, Lippi, and Zaffaroni}{Forni
  et~al.}{2017}]{FHLZ15}
Forni, M., M.~Hallin, M.~Lippi, and P.~Zaffaroni (2017).
\newblock Dynamic factor models with infinite dimensional factor space:
  Asymptotic analysis.
\newblock {\em Journal of Econometrics\/}~{\em 199}, 74--92.

\bibitem[\protect\citeauthoryear{Forni and Lippi}{Forni and
  Lippi}{2001}]{fornilippi01}
Forni, M. and M.~Lippi (2001).
\newblock The {G}eneralized {D}ynamic {F}actor {M}odel: {R}epresentation
  theory.
\newblock {\em Econometric Theory\/}~{\em 17}, 1113--1141.

\bibitem[\protect\citeauthoryear{Geweke}{Geweke}{1977}]{geweke77}
Geweke, J. (1977).
\newblock The dynamic factor analysis of economic time series.
\newblock In D.~J. Aigner and A.~S. Goldberger (Eds.), {\em Latent Variables in
  Socio-Economic Models}. Amsterdam: North Holland.

\bibitem[\protect\citeauthoryear{Hallin and Lippi}{Hallin and
  Lippi}{2013}]{HALLIPPI13}
Hallin, M. and M.~Lippi (2013).
\newblock Factor models in high-dimensional time series--{A} time-domain
  approach.
\newblock {\em Stochastic Processes and their Applications\/}~{\em 123\/}(7),
  2678--2695.

\bibitem[\protect\citeauthoryear{Hallin, Lippi, Barigozzi, Forni, and
  Zaffaroni}{Hallin et~al.}{2020}]{hallin20}
Hallin, M., M.~Lippi, M.~Barigozzi, M.~Forni, and P.~Zaffaroni (2020).
\newblock {\em Time Series in High Dimensions: The General Dynamic Factor
  Model}.
\newblock Singapore: World {Scientific}.

\bibitem[\protect\citeauthoryear{Hallin and Li{\v s}ka}{Hallin and Li{\v
  s}ka}{2007}]{hallinliska07}
Hallin, M. and R.~Li{\v s}ka (2007).
\newblock Determining the number of factors in the general dynamic factor
  model.
\newblock {\em Journal of the American Statistical Association\/}~{\em 102},
  603--617.

\bibitem[\protect\citeauthoryear{Hannan and Deistler}{Hannan and
  Deistler}{2012}]{deistler12}
Hannan, E.~J. and M.~Deistler (2012).
\newblock {\em The statistical theory of linear systems}.
\newblock Philadelphia: SIAM Edition, Republication of the work first published
  by John Wiley and Sons, in 1988.

\bibitem[\protect\citeauthoryear{Onatski}{Onatski}{2012}]{onatskiweak}
Onatski, A. (2012).
\newblock {Asymptotics of the principal components estimator of large factor
  models with weakly influential factors}.
\newblock {\em Journal of Econometrics\/}~{\em 168\/}(2), 244--258.

\bibitem[\protect\citeauthoryear{Poncela, Ruiz, and Miranda}{Poncela
  et~al.}{2021}]{PONCELA2021}
Poncela, P., E.~Ruiz, and K.~Miranda (2021).
\newblock Factor extraction using kalman filter and smoothing: This is not just
  another survey.
\newblock {\em International Journal of Forecasting\/}~{\em 37\/}(4),
  1399--1425.

\bibitem[\protect\citeauthoryear{Quah and Sargent}{Quah and
  Sargent}{1993}]{QuahSargent93}
Quah, D. and T.~J. Sargent (1993, December).
\newblock {A Dynamic Index Model for Large Cross Sections}.
\newblock In {\em {Business Cycles, Indicators, and Forecasting}}, NBER
  Chapters, pp.\  285--310. National Bureau of Economic Research, Inc.

\bibitem[\protect\citeauthoryear{Sargent and Sims}{Sargent and
  Sims}{1977}]{SargentSims}
Sargent, T.~J. and C.~A. Sims (1977).
\newblock Business cycle modeling without pretending to have too much a priori
  economic theory.
\newblock In C.~A. Sims (Ed.), {\em New methods in business cycle research}.
  Federal Reserve Bank of Minneapolis.

\bibitem[\protect\citeauthoryear{Scherrer and Deistler}{Scherrer and
  Deistler}{1998}]{scherdeist98}
Scherrer, W. and M.~Deistler (1998).
\newblock A structure theory for linear dynamic errors-in-variables models.
\newblock {\em SIAM Journal on Control and Optimization\/}~{\em 36},
  2148--2175.

\bibitem[\protect\citeauthoryear{Shumway and Stoffer}{Shumway and
  Stoffer}{2000}]{shumway2000}
Shumway, R. and D.~Stoffer (2000).
\newblock {\em Time series Analysis and its applications}.
\newblock New York: Springer.

\bibitem[\protect\citeauthoryear{Spearman}{Spearman}{1904}]{spearman04}
Spearman, C. (1904).
\newblock "{General} intelligence," objectively determined and measured.
\newblock {\em The American Journal of Psychology\/}~{\em 15\/}(2), 201--292.

\bibitem[\protect\citeauthoryear{Stock and Watson}{Stock and
  Watson}{2002a}]{stockwatson02JASA}
Stock, J.~H. and M.~W. Watson (2002a).
\newblock Forecasting using principal components from a large number of
  predictors.
\newblock {\em Journal of the American Statistical Association\/}~{\em 97},
  1167--1179.

\bibitem[\protect\citeauthoryear{Stock and Watson}{Stock and
  Watson}{2002b}]{stockwatson02JBES}
Stock, J.~H. and M.~W. Watson (2002b).
\newblock Macroeconomic forecasting using diffusion indexes.
\newblock {\em Journal of Business and Economic Statistics\/}~{\em 20},
  147--162.

\bibitem[\protect\citeauthoryear{Stock and Watson}{Stock and
  Watson}{2016}]{stockwatson16}
Stock, J.~H. and M.~W. Watson (2016).
\newblock Dynamic factor models, factor-augmented vector autoregressions, and
  structural vector autoregressions in macroeconomics.
\newblock In J.~B. Taylor and H.~Uhlig (Eds.), {\em Handbook of
  Macroeconomics}, Volume~2, pp.\  415--525. Amsterdam: Elsevier.

\bibitem[\protect\citeauthoryear{Watson and Engle}{Watson and
  Engle}{1983}]{WatsonEngle83}
Watson, M.~W. and R.~F. Engle (1983).
\newblock Alternative algorithms for the estimation of dynamic factor, mimic
  and varying coefficients regression models.
\newblock {\em Journal of Econometrics\/}~{\em 23}, 385--400.

\end{thebibliography}
\end{document}